%% file: main.tex
\documentclass{article}

\usepackage[final]{neurips_2024}

\usepackage[utf8]{inputenc} 
\usepackage[T1]{fontenc}    
\usepackage{hyperref}       
\usepackage{url}            
\usepackage{booktabs}       
\usepackage{amsfonts}       
\usepackage{nicefrac}       
\usepackage{microtype}      
\usepackage{xcolor}         

\usepackage{graphicx}
\usepackage{mathtools}
\usepackage{algorithm}
\usepackage[noend]{algpseudocode} 
\usepackage{colortbl}
\usepackage{amsmath}
\usepackage{amsthm}
\usepackage{amssymb}
\usepackage[capitalize,noabbrev]{cleveref}
\usepackage{tikz}
\usepackage{xspace}
\usepackage{graphicx}
\usepackage{wrapfig}
\usepackage{caption}
\usepackage{subcaption}

\newcommand{\ie}{\textit{i.e.,}\xspace}

\newcommand{\Csample}{$a_c$\xspace}

\newcommand{\worstsample}{$a_w$\xspace}

\newtheorem{theorem}{Theorem}

\newtheorem{corollary}[theorem]{Corollary}
\theoremstyle{definition}

\theoremstyle{remark}

\title{Adversarial Sample-Based Approach for Tighter Privacy Auditing in Final Model-Only Scenarios}

\author{%
  Sangyeon Yoon\thanks{These authors contributed equally.} \\
  Hongik University\\
  \texttt{B911117@g.hongik.ac.kr} \\
  \And
  Wonje Jeung\footnotemark[1]\\
  Yonsei University\\
  \texttt{specific0924@yonsei.ac.kr} \\
  \AND
  Albert No\\
  Yonsei University\\
  \texttt{albertno@yonsei.ac.kr} 
}

\begin{document}

\maketitle

\begin{abstract}
Auditing Differentially Private Stochastic Gradient Descent (DP-SGD) in the final model setting is challenging and often results in empirical lower bounds that are significantly looser than theoretical privacy guarantees. We introduce a novel auditing method that achieves tighter empirical lower bounds without additional assumptions by crafting worst-case adversarial samples through loss-based input-space auditing. Our approach surpasses traditional canary-based heuristics and is effective in final model-only scenarios. Specifically, with a theoretical privacy budget of $\varepsilon = 10.0$, our method achieves empirical lower bounds of $4.914$, compared to the baseline of $4.385$ for MNIST. 
Our work offers a practical framework for reliable and accurate privacy auditing in differentially private machine learning.
\end{abstract}

\section{Introduction} \label{sec:intro}

Differentially Private Stochastic Gradient Descent (DP-SGD)~\citep{abadi2016deep} was introduced to prevent sensitive information leakage from training data in models trained using Stochastic Gradient Descent (SGD)~\citep{shokri2017membership,hayes2017logan,yeom2018privacy,bichsel2021dp,balle2022reconstructing,haim2022reconstructing,carlini2022membership}. DP-SGD mitigates privacy leakage by clipping individual gradients and adding Gaussian noise to the aggregated gradient updates.

However, correctly implementing DP-SGD is challenging, and several implementations have revealed bugs that compromise its privacy guarantees~\citep{ding2018detecting,bichsel2021dp,stadler2022synthetic,tramer2022debugging}. These issues introduce potential privacy leakage, making thorough audits essential to ensure that privacy guarantees hold in practice. Audits commonly employ membership inference attacks (MIAs)~\citep{shokri2017membership}, using success rates to empirically estimate privacy leakage and compare it against theoretical upper bounds~\citep{jagielski2020auditing,nasr2021adversary,nasr2023tight}. If the empirical privacy leakage, represented by the lower bound on $\varepsilon$, is significantly lower than the theoretical upper bound, the audit results are considered \emph{loose}; conversely, when the lower bound closely approaches the upper bound, the results are regarded as \emph{tight}.
\begin{figure*}
    \centering
    \includegraphics[width=\textwidth]{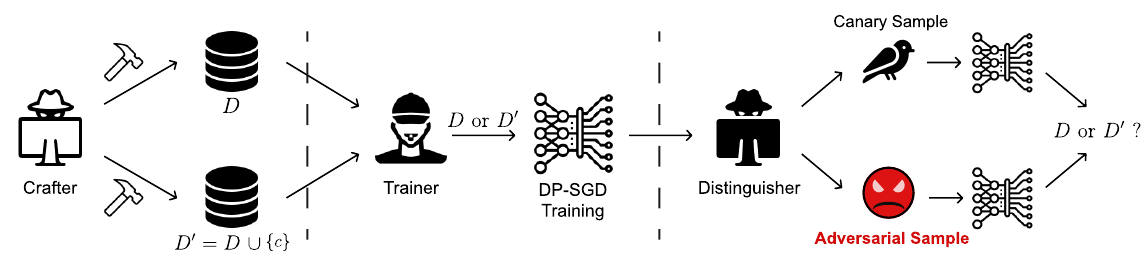}
    \caption{Overview of privacy auditing process. The Crafter crafts a canary and constructs two neighboring datasets, $D$ and $D'$, where $D'$ contains the canary.
    The Trainer trains a model on either $D$ or $D'$ using the DP-SGD.
    The Distinguisher observes the loss values for specific input to infer whether the model was trained on $D$ or $D'$.
    We suggests an adversarial sample for \textit{tighter} auditing.
    }
    \label{fig:overview}
\end{figure*}
In \emph{intermediate model} settings, where adversaries can observe gradients throughout the entire training process, previous works~\citep{nasr2021adversary,nasr2023tight} have shown that achieving tight privacy auditing is feasible. However, this approach is often impractical, as real-world scenarios usually restrict adversaries' access to only the final model. Achieving rigorous audit results in the \emph{final model} settings remains challenging, leaving the feasibility of final model auditing for DP-SGD an unresolved issue.

To address these challenges, the final model setting has gained attention in practical applications as a promising method to improve the privacy-utility trade-off. By concealing intermediate steps and exposing only the final model, this approach aligns more closely with real-world constraints. Theoretically, in the final model setting, privacy amplification by iteration can provide additional privacy protection for data points involved in earlier stages of training~\citep{feldman2018privacy,balle2019privacy,chourasia2021differential,ye2022differentially,altschuler2022privacy,bok2024shifted}. However, this effect has primarily been observed in convex or smooth loss functions, leaving its applicability to non-convex problems unresolved.

Building on this theoretical foundation, several works~\citep{nasr2021adversary,nasr2023tight,andrew2023one,steinke2024privacy,steinke2024last,cebere2024tighter} have explored privacy amplification in final model settings with non-convex loss functions. They observed a significant gap between empirical lower bounds and theoretical upper bounds, suggesting that privacy amplification by iteration may extend to general non-convex loss functions. However, there is no formal proof to support this, and the observed gap may instead reflect that adversaries have not fully exploited their potential capabilities.

Conversely, recent works~\citep{annamalai2024s,cebere2024tighter} have observed that privacy amplification does not occur in non-convex settings. These findings, however, apply only to specific cases, such as constructing a worst-case non-convex loss function for DP-SGD where information from all previous iterations is encoded in the final iteration, or manually designing a gradient sequence. These efforts remain constrained to additional assumptions or highly specific scenarios, leaving the broader question of whether privacy amplification can occur in general non-convex settings unresolved.

In this work, we introduce a novel auditing approach designed to establish tighter lower bounds in general non-convex settings for final model scenarios, without requiring any additional assumptions.
Our method leverages loss-based input-space auditing to craft worst-case adversarial samples,
enhancing the precision of privacy auditing results.
Previous works~\citep{jagielski2020auditing,nasr2021adversary,tramer2022debugging,aerni2024evaluations} on input-space auditing 
have implicitly assumed that the canary sample represents the worst-case privacy leakage,
using it as an adversarial sample for MIAs based on its loss outputs.
However, relying solely on canary-based approaches may not yield tighter empirical lower bounds.
By leveraging loss outputs from an alternative adversarial sample, our approach identifies increased privacy leakage.
Consequently, we craft worst-case adversarial samples without relying on canaries,
providing tighter empirical lower bounds for auditing.

Our auditing method addresses practical challenges in the final model setting.
Adversaries have access to the final model's weights, as in open-source cases. We craft the worst-case adversarial sample directly using these weights. For example, with a theoretical privacy budget of $10.0$, our approach achieves an empirical lower bound of $4.914$, significantly surpassing the $4.385$ obtained with the canary approach.

\section{Preliminary}

\paragraph{Differentially Private Training.}
Differential Privacy (DP)~\citep{dwork2006calibrating} is a widely used standard for preserving individual privacy in data analysis.
A randomized mechanism $\mathcal{M}$ is considered $(\varepsilon, \delta)$-differentially private if, 
for any two neighboring datasets $D$ and $D'$ that differ by only one data point known as the \textbf{canary} ($c$), 
and for any possible output subset $O \subseteq \textit{Range}(\mathcal{M})$, the following inequality holds:
\begin{equation*}
    \mathrm{Pr}[\mathcal{M}(D) \in O] \leq e^\varepsilon\, \mathrm{Pr}[\mathcal{M}(D') \in O] + \delta
\end{equation*}
In this formulation, $\varepsilon$ represents the loss of privacy, with smaller $\varepsilon$ indicating stronger privacy guarantees. 
The parameter $\delta \in (0,1)$ represents a small probability that the privacy guarantee may not hold.

Numerous mechanisms have been developed to ensure differential privacy during the training of machine learning models.
~\citep{abadi2016deep,papernot2016semi,zhu2020private}.
Differentially Private Stochastic Gradient Descent (DP-SGD)~\citep{abadi2016deep} is widely used to achieve differential privacy.
DP-SGD clips gradients to a maximum norm $C$ to limit the impact of individual data points on model parameters, 
then adds Gaussian noise scaled by $\sigma$ to ensure privacy across training steps.
This mechanism allows DP-SGD to provide a theoretical upper bound on cumulative privacy loss, 
ensuring privacy throughout training (known as \textbf{privacy accounting}).
The complete DP-SGD algorithm is shown in~\cref{alg:dp_sgd}.

\input{algorithms/dp_sgd}

\paragraph{Privacy Auditing.} 
Although DP-SGD provides a theoretical upper bound on privacy budgets ($\varepsilon$, $\delta$), 
relying solely on privacy accounting poses challenges.
First, Theoretical analyses, while crucial, are demonstrated to be conservative
~\citep{bassily2014private,kairouz2015composition,abadi2016deep,mironov2017renyi,koskela2020computing,gopi2021numerical,doroshenko2022connect}, 
which can lead to overestimating the required noise and subsequently reducing the utility of the model~\citep{nasr2021adversary}.
Second, the complexity of DP-SGD training can lead to implementation errors that compromise privacy guarantees~\citep{ding2018detecting,tramer2022debugging}.
Privacy auditing addresses these issues by establishing empirical lower bounds on privacy, 
providing a more realistic assessment of privacy loss.

In privacy auditing,~\citet{nasr2021adversary} decomposes the attack process into three main components: \textbf{Crafter}, \textbf{Trainer}, and \textbf{Distinguisher}.
The Crafter generates a canary sample $c$ to distinguish between neighboring datasets $D$ and $D'$ (\ie $D' = D \cup \{c\}$).
The Trainer then trains a model on one of these two datasets using the DP-SGD training algorithm.
Finally, the Distinguisher receives the datasets $D$ and $D'$ from the Crafter, along with the trained model as input.
The distinguisher then infers which dataset was used,
with the accuracy of this inference indicating the level of privacy leakage.
Together, the Crafter and the Distinguisher form the \textbf{adversary} ($\mathcal{A}$).

\paragraph{Privacy Auditing Setup.}
Privacy auditing can be categorized into two settings based on adversaries' access levels:
intermediate models~\citep{nasr2021adversary,nasr2023tight,andrew2023one, steinke2024privacy,mahloujifar2024auditing}, 
final model~\citep{jagielski2020auditing,nasr2021adversary,annamalai2024nearly,cebere2024tighter,steinke2024last}.
For the \textbf{intermediate models} setup,
the adversaries have full access to the model parameters during all training steps.
Previous works~\citep{nasr2021adversary,nasr2023tight} have demonstrated that privacy auditing in this setup achieves a \textit{tight} lower bound, where the theoretical upper bound and the empirical lower bound ($\varepsilon_{emp}$) can align.

However, in practice, adversaries are more likely to have access only to the \textbf{final model} rather than to all intermediate training steps.
Access to the final model is available either through its outputs, such as via APIs, or through direct access to its weights, as in open-source models.
In the case of access to the final model through an API, a substantial gap remains between $\varepsilon$ and $\varepsilon_{emp}$.
Although \citet{annamalai2024nearly} attempts to address this, 
their approach requires additional assumptions on initialized parameters 
and still lacks tightness.
In contrast, recent works~\citep{annamalai2024s,cebere2024tighter} aim for tight auditing via direct access to model weights in the final model setting. 
However, their auditing methods focus on specific scenarios, such as constructing worst-case non-convex loss functions for DP-SGD, manually designing gradient sequences, or assuming access to the initial parameters ($\theta_0$).
Our work delivers tighter privacy auditing \textbf{without additional assumptions}, relying solely on final model.

Privacy auditing techniques can be broadly classified into gradient-space auditing and input-space auditing, 
based on the modifications allowed by the Crafter. 
In gradient-space auditing~\citep{nasr2021adversary,maddock2022canife,nasr2023tight,andrew2023one,steinke2024privacy}, the gradient canary is embedded directly into the gradients during training, allowing targeted gradient-level interventions. 
In contrast, 
input space auditing~\citep{jagielski2020auditing,zanella2023bayesian,andrew2023one,steinke2024last} uses a canary in the form of an input sample,
focusing on the training data itself.
In our experiments, we adopt input-space auditing, 
which is more practical.

\input{algorithms/our-auditing}
\section{DP-SGD Auditing procedure}
Recent privacy auditing studies~\citep{nasr2021adversary, tramer2022debugging, nasr2023tight, steinke2024last, cebere2024tighter, chadha2024auditing} often employ membership inference attacks (MIAs)~\citep{shokri2017membership} to determine whether a specific sample was included in the training data by analyzing the outputs of $\mathcal{M}(D)$ or $\mathcal{M}(D')$. Similarly, our auditing procedure in~\cref{alg:auditing} uses MIAs to evaluate the privacy guarantees of the model, consistent with their widespread adoption as the de facto standard in many privacy studies~\citep{jeon2024information,thaker2024position}.

The procedure begins with Crafter phase.
A canary Crafter crafts a canary sample $c$ as an outlier within the training dataset $D$.
This canary sample is then inserted into the original dataset to create a neighboring dataset,
$D' = D \cup \{c\}$.
In our experiments, following prior works~\citep{de2022unlocking, nasr2023tight, steinke2024last}, we primarily use a blank image as the canary to maximize its distinctiveness from typical training samples.

In the \textbf{Model Trainer} phase, trainer applies the DP-SGD mechanism $\mathcal{M}$ to both $D$ and $D'$, producing $N$ models for each dataset, denoted as $\{M_i\}_{i=1}^N$ and $\{M'_i\}_{i=1}^N$, respectively. 
These models are trained independently to analyze the effect of canary insertion across multiple models.

The \textbf{Distinguisher} phase performs the core auditing process. 
It first crafts an adversarial sample $a$, 
then computes the loss values of this sample across all models, forming output sets $O$ and $O'$ for $\{M_i\}_{i=1}^N$ and $\{M'_i\}_{i=1}^N$, respectively. 
Using a decision threshold $\tau$ determined by a separate dataset (e.g., a validation set), 
the Distinguisher calculates the False Positive Rate (FPR) and False Negative Rate (FNR) based on $O$ and $O'$. 
It then computes the upper bounds of FPR and FNR using Clopper-Pearson confidence intervals~\citep{clopper1934use}. 
Finally, based on the empirical upper bounds $\overline{\text{FPR}}$ and $\overline{\text{FNR}}$, the Distinguisher estimates the empirical lower bound for the privacy parameter $\varepsilon$ at a given $\delta$, denoted as $\varepsilon_{\text{emp}}$.

\citet{nasr2023tight} demonstrated that the privacy region for DP-SGD aligns more closely with the $\mu$-Gaussian Differential Privacy guarantee ($\mu$-GDP)~\citep{dong2022gaussian} than with the $(\varepsilon, \delta)$-DP guarantee, establishing a direct relationship between $\overline{\text{FPR}}$, $\overline{\text{FNR}}$, and $\mu$.
Furthermore, it enables a tighter estimate of privacy leakage with fewer training runs than $(\varepsilon, \delta)$-DP.
In addition, when deciding the decision threshold, any value for $\mu$-GDP
has an equal likelihood of maximizing the lower bound given a sufficient number of observations~\citep{nasr2023tight}.
Following previous work~\citep{nasr2021adversary,maddock2022canife,zanella2023bayesian}, we use the threshold that maximizes the $\varepsilon$ lower bound for the same set of observations.

Calculating a empirical bound for $\mu$ in $\mu$-GDP, which can be converted to a lower bound for $\varepsilon$ in $(\varepsilon, \delta)$-DP, provides an effective estimate of privacy leakage.
To compute a lower bound on the privacy parameters of the Gaussian mechanism (i.e., \( \mu \)), we have:
\begin{equation*}
    {\mu}_{emp} = \Phi^{-1}(1 - \overline{\text{FPR}}) - \Phi^{-1}(\overline{\text{FNR}}),
\end{equation*}
where $\Phi^{-1}$ is the inverse of the standard normal CDF.
Then, ${\mu}_{emp}$ corresponds to ${\varepsilon}_{emp}$ by following Corollary.

\begin{corollary}[\( \mu \)-GDP to \((\varepsilon, \delta)\)-DP conversion~\citep{dong2022gaussian}] A mechanism is \( \mu \)-GDP 
if and only if it is \((\varepsilon, \delta(\varepsilon))\)-DP for all \( \varepsilon \geq 0 \), where:
\begin{equation*}
    \delta(\varepsilon) = \Phi \left( -\frac{\varepsilon}{\mu} + \frac{\mu}{2} \right) - e^{\varepsilon} \Phi \left( -\frac{\varepsilon}{\mu} - \frac{\mu}{2} \right).
\end{equation*}
$\Phi$ is the standard normal CDF.
\end{corollary}
We apply this $\mu$-GDP based auditing scheme with $\delta = 10^{-5}$ across all experiments.

\begin{figure*}
    \centering
    \begin{subfigure}[b]{0.24\linewidth} 
        \centering
        \includegraphics[width=\linewidth]{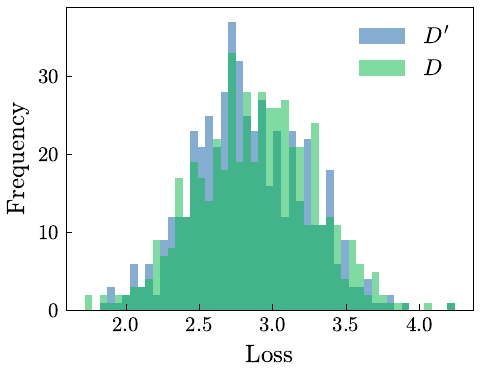}
        \caption{\Csample}
        \label{fig:loss_results_canary}
    \end{subfigure}
    \begin{subfigure}[b]{0.24\linewidth} 
        \centering
        \includegraphics[width=\linewidth]{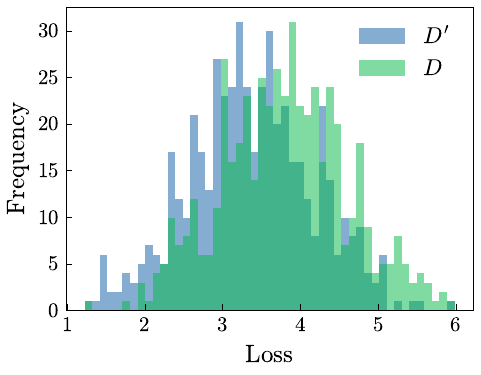}
        \caption{$a_{L_2}$}
        \label{fig:loss_results_UDE}
    \end{subfigure}
    \begin{subfigure}[b]{0.24\linewidth} 
        \centering
        \includegraphics[width=\linewidth]{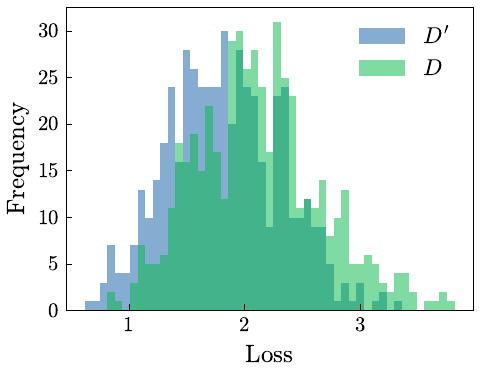}
        \caption{$a_{F}$}
        \label{fig:loss_results_UDE}
    \end{subfigure}
    \begin{subfigure}[b]{0.24\linewidth} 
        \centering
        \includegraphics[width=\linewidth]{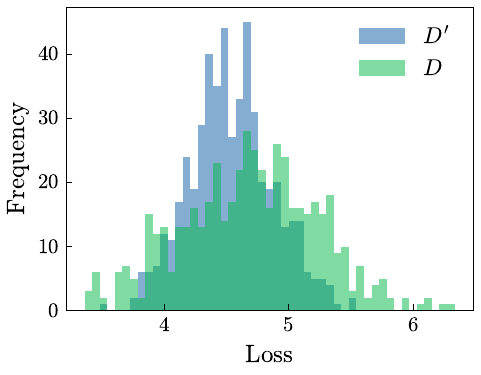}
        \caption{$a_{BD}$}
        \label{fig:loss_results_UDE}
    \end{subfigure}
    \caption{
    The loss distributions of the canary sample \( a_c \) and three adversarial samples \( (a_{L_2}, a_{F}, a_{BD}) \) in the training set at a fixed privacy budget \( \varepsilon = 1.0 \).
    The green distribution represents the loss outputs of models \(M\), while the blue distribution represents the loss outputs of models \(M'\).
    }
    \label{fig:loss_results}
\end{figure*}
\section{Crafting Adversarial Samples for Tighter Privacy Auditing}

To obtain tighter lower bounds for membership inference attacks (MIAs), some studies~\citep{nasr2021adversary,nasr2023tight} craft input canaries based on the final model weights during line 3 of \cref{alg:auditing}.
In contrast, most prior works~\citep{carlini2019secret,jagielski2020auditing,nasr2021adversary,aerni2024evaluations,annamalai2024nearly} rely on the heuristic assumption that the most vulnerable data point serves as the canary sample in the line 8 of the same algorithm.
In our study, instead of relying on a canary-based approach, we directly construct the worst-case adversarial sample,
adhering to the requirement that every output must satisfy the differential privacy guarantee,
thereby ensuring fundamental indistinguishability across all possible input samples.
To the best of our knowledge, this is the first attempt in loss-based input-space auditing to achieve tighter lower bounds by using a worst-case adversarial sample instead of a traditional canary.

To enable tighter privacy auditing, we craft a worst-case adversarial sample $a_w$ that maximizes the distinguishable difference between the distributions of outputs $O$ and $O'$. This involves designing a specialized loss function to enhance the separability of these distributions.
When crafting $a_w$, we leverage the Distinguisher's knowledge of the canary 
by initializing the adversarial sample $a=(x_a, y_a)$ with the canary's values $c=(x_c, y_c)$.
Each pixel of $x_a$ is treated as an learnable variable to form the worst-case scenario,
while the label $y_a$ is fixed as a constant $y_c$.

\paragraph{Distance-Based Separation.}
Intuitively, the canary's loss tends to be higher for models trained on the dataset \(D\) (denoted by \(M_i\)) than for those trained on \(D'\) (denoted by \(M'_i\)). To exploit this discrepancy, we train an adversarial sample \(a_w\) that encourages \(M_i\) to produce a higher loss and \(M'_i\) to produce a lower loss, thereby maximizing the distinguishability between their output distributions. 
As a straightforward approach, we utilize the L2 loss, \(\mathcal{L}_{L_2}\), to widen this gap:
\[
    \mathcal{L}_{L_2}(x_a) 
    = \frac{1}{N}\sum_{i=1}^{N} \Bigl[ \ell\bigl(M'_i(x_a), y_a\bigr) \;-\; \ell\bigl(M_i(x_a), y_a\bigr) \Bigr],
\]
and denote the adversarial sample generated by this objective as \(a_{L_2}\).


\paragraph{Distributional Separation.}  
Additionally, we apply two distribution-based losses to maximize the separation between \(M'\) and \(M\). 
The first is the \emph{Bhattacharyya loss}, \(\mathcal{L}_{BD}\), which is derived from the Bhattacharyya distance, a measure of overlap between two probability distributions. 
When the outputs of \(M_i(x_a)\) and \(M'_i(x_a)\) can be approximated as univariate Gaussians 
\(\mathcal{N}(\mu_{M_i}, \sigma_{M_i}^2)\) and 
\(\mathcal{N}(\mu_{M'_i}, \sigma_{M'_i}^2)\), 
the Bhattacharyya distance between them is:
\[
D_B\bigl(\mathcal{N}(\mu_{M_i}, \sigma_{M_i}^2), 
         \mathcal{N}(\mu_{M'_i}, \sigma_{M'_i}^2)\bigr)
= \frac{(\mu_{M_i} - \mu_{M'_i})^2}
       {4\bigl(\sigma_{M_i}^2 + \sigma_{M'_i}^2\bigr)}
\;+\;\frac12 \ln\!\Bigl(
      \frac{\sigma_{M_i}^2 + \sigma_{M'_i}^2}
           {2\,\sigma_{M_i}\,\sigma_{M'_i}} \Bigr).
\]
To minimize this overlap, we define \(\mathcal{L}_{BD}\) as the negative of the Bhattacharyya distance:
\[
\mathcal{L}_{BD} = 
    - \frac{1}{N}\sum_{i=1}^{N} 
      D_B\bigl(\mathcal{N}(\mu_{M_i}, \sigma_{M_i}^2), 
               \mathcal{N}(\mu_{M'_i}, \sigma_{M'_i}^2)\bigr).
\]
Minimizing \(\mathcal{L}_{BD}\) thereby pushes the two distributions apart, enhancing the distinguishability between \(M\) and \(M'\).

Additionally we adopt
\emph{Fisher's Linear Discriminant Analysis (LDA)}. 
LDA maximizes the ratio of the squared difference between class means to the sum of their variances, thereby encouraging a clear separation between distributions. 
Following this principle, we define the \emph{Fisher Discriminant Loss} as:
\[
    \mathcal{L}_{F} 
    = -\frac{1}{N} \sum_{i=1}^{N} 
      \frac{(\mu_{M'_i} - \mu_{M_i})^2}
           {\sigma_{M'_i}^2 + \sigma_{M_i}^2}.
\]
By normalizing the squared difference of the means by the sum of their variances, \(\mathcal{L}_{F}\) remains stable and effectively preserves the separation between \(M\) and \(M'\). 
We denote the adversarial samples produced by minimizing the Bhattacharyya loss and the Fisher Discriminant Loss as \(a_{BD}\) and \(a_F\), respectively; both aim to maximize the distributional gap between \(M\) and \(M'\).



\paragraph{Data Splitting for Adversarial Samples.}
To mitigate the risk of information leakage when generating adversarial samples, we adopt a strict data-splitting protocol. Specifically, we derive the adversarial samples exclusively from the training models and then measure the empirical lower bounds of differential privacy on an independent validation models. This separation ensures that any potential leakage from the training process does not influence the subsequent evaluation, thereby yielding a more reliable assessment of the privacy guarantees.  Overall, this strategy effectively decouples the adversarial sample generation from the empirical DP evaluation, reducing bias and strengthening the credibility of our results.

\section{Experiments}

\subsection{Experiments Setup}
To evaluate our auditing procedure, we conduct experiments on the MNIST~\citep{lecun1998mnist} dataset using ConvNet models. Following the architecture used in~\citet{annamalai2024nearly}, we train the models with a learning rate of \( \eta = 4 \) for \( T = 100 \) iterations. For models trained with DP-SGD, the accuracy on MNIST remains consistently close to 95\% across all values of \( \varepsilon \), indicating the stability of the training process under differential privacy constraints.

\paragraph{Implementation Details.}
To ensure robustness, we perform 10 independent runs for each experiment and report the average value of $\varepsilon_{\text{emp}}$ along with its standard deviation, 
and the theoretical upper bound for $\varepsilon$ is computed using the privacy accountant provided by Opacus~\citep{yousefpour2021opacus}.
In each run, we generate $2N = 1024$ models--512 with canary ($M'$) and 512 without canary ($M$).
We compute empirical lower bounds using the $\mu$-GDP approach, following previous works~\citep{nasr2023tight,cebere2024tighter}, and report the lower bounds with 95\% confidence intervals.

Additionally, to simplify the privacy analysis, we set the sampling rate to 1, performing DP-SGD with full-batch training ($B = |D|$). 
While, analyzing DP-SGD with both subsampling amplification and multi-step composition is challenging, 
and relying on the central limit theorem for subsampling may underestimate the privacy budget due to the emergence of mixture distributions~\citep{nasr2023tight}. 
Although the Privacy Loss Distribution (PLD)~\citep{koskela2020computing} is used to approximate the trade-off function and address this issue~\citep{nasr2023tight}, 
PLD lacks a closed-form trade-off function and provides a looser bound for $\varepsilon$.
For these reasons, we focus our analysis on the full-batch setting.

\begin{wraptable}{r}{7cm}
\vskip -0.3in
\caption{
Privacy auditing results on the MNIST dataset with various adversarial samples (Canary, L2, Fisher, Bhattacharyya) in theoretical privacy budget of $\varepsilon = \{1.0,2.0,4.0,10.0\}$. }
    \label{tab:various_alpha}
 \centering
    \resizebox{7cm}{!}{%
    \begin{tabular}{c|cccc}
        \midrule
        \cmidrule(lr){1-1} \cmidrule(lr){2-5} 
        Loss & $\varepsilon=1.0$ & $\varepsilon=2.0$ & $\varepsilon=4.0$ & $\varepsilon=10.0$ \\ 
        \midrule
        Canary
        & $0.212_{\pm 0.177}$
        & $0.594_{\pm 0.107}$
        & $1.696_{\pm 0.207}$
        & $4.385_{\pm 0.284}$\\
        L2-Distance
        & $0.215_{\pm 0.119}$
        & $0.646_{\pm 0.142}$
        & $1.722_{\pm 0.248}$
        & $4.788_{\pm 0.255}$\\
        Fisher
        & $\underline{0.226}_{\pm 0.112}$
        & $\underline{0.648}_{\pm 0.116}$
        & $\textbf{1.823}_{\pm 0.335}$
        & $\textbf{4.914}_{\pm 0.268}$\\
        Bhattacharyya
        & $\textbf{0.232}_{\pm 0.141}$
        & $\textbf{0.661}_{\pm 0.120}$
        & $\underline{1.785}_{\pm 0.297}$
        & $\underline{4.854}_{\pm 0.268}$\\
        \midrule
    \end{tabular} 
    }
\end{wraptable}
\subsection{Auditing Results in Final Model}
\label{sec:auditing_results}

To evaluate the effectiveness of adversarial sample selection in privacy auditing, we compare the empirical privacy lower bounds \( \varepsilon_{\text{emp}} \) obtained using the default canary sample (\Csample) and our crafted worst-case adversarial samples (\worstsample) in~\cref{tab:various_alpha}. Specifically, \Csample\ refers to the default canary that differentiates \( D \) and \( D' \), whereas \worstsample\ consists of adversarial samples generated using the loss functions \( \mathcal{L}_{L_2} \), \( \mathcal{L}_{BD} \), and \( \mathcal{L}_{F} \). 

\begin{wrapfigure}{r}{0.37\textwidth}
\vskip -.18in
    \centering
        \includegraphics[width=0.37\textwidth]{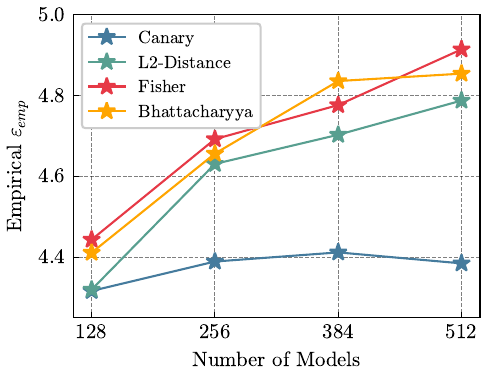}
    \caption{Privacy auditing results on the MNIST dataset using $\{128, 256, 384, 512\}$ models.}
    \label{fig:num_models}
\vskip -.5in
\end{wrapfigure}
Across different theoretical privacy budgets \( \varepsilon \), employing \worstsample\ consistently yields tighter empirical lower bounds than the baseline \Csample. As shown in~\cref{tab:various_alpha}, for \( \varepsilon = 1.0 \), the default canary achieves an empirical bound of 0.212, while Fisher and Bhattacharyya loss-based samples improve it to 0.226 and 0.232, respectively. This trend persists at higher privacy budgets, where Fisher-based samples achieve the highest empirical bound of 4.914 at \( \varepsilon = 10.0 \), outperforming the canary baseline (4.385). Notably, both Fisher and Bhattacharyya-based adversarial samples consistently provide the tightest privacy lower bounds across all settings, confirming their effectiveness in maximizing privacy distinguishability.

\subsection{Effect of Number of Models}
As shown in \Cref{fig:num_models}, adversarial samples consistently achieve higher auditing performance as the number of models grows, rising from 4.44 with 128 models to 4.91 with 512 models for $a_F$. 
This improvement arises because, as the number of models grows, the distributions of training-model and evaluation-model losses become more similar. This closer alignment enables adversarial samples to be more effectively optimized, ultimately leading to enhanced auditing outcomes.

\section{Conclusion}
We propose a novel method for crafting worst-case adversarial samples that tighten lower bounds in privacy auditing of differentially private models, focusing on the final model setting without additional assumptions. 
In contrast to traditional canary-based approaches, our method generates adversarial samples that maximize the distinguishability between output distributions. 
Experiments on the MNIST dataset demonstrate that this approach provides substantially tighter empirical lower bounds on privacy leakage compared to baseline methods. 
These findings highlight the promise of employing specialized adversarial samples for more rigorous privacy auditing, and pave the way for extending such techniques to more complex datasets and model architectures.

\bibliographystyle{plainnat}
\bibliography{main}

\end{document}

%% file: algorithms/dp_sgd.tex
\begin{algorithm}[H] 
    \small
    \caption{DP-SGD}
    \label{alg:dp_sgd}
    \begin{algorithmic}[1]
    \State \textbf{Input:} Training dataset $D$, Loss function $\ell$, Iterations $T$, Initial model parameters $\theta_0$,  Learning rate $\eta$, Batch size $B$, Clipping norm $C$, Noise multiplier $\sigma$
    \For{$t \in [T]$}
        \State Sample $L_t$ from $D$ using poisson sampling with 
        \State probability $B/|D|$
        \For{$x_i \in L_t$}
            \State $g_t(x_i) \leftarrow \nabla_{\theta_t} \ell(\theta_t; x_i)$
            \State $\bar{g_t}(x_i) \leftarrow g_t(x_i) / \max(1, \frac{||g_t(x_i)||_2}{C})$
        \EndFor
        \State $\tilde{g}_t \leftarrow \frac{1}{B} \left(\sum_i \bar{g_t}(x_i) + \mathcal{N}(0, (C\sigma )^2 \mathbb{I})\right)$
        \State $\theta_{t + 1} \leftarrow \theta_t - \eta \tilde{g}_t$
    \EndFor
    \State \Return $\theta_T$ 
    \end{algorithmic} 
\end{algorithm}

%% file: algorithms/our-auditing.tex
\begin{algorithm}[H]
    \small
    \caption{Auditing Procedure}\label{alg:audit_dpsgd}
    \begin{algorithmic}[1]
    \State \textbf{Input:} Training dataset $D$, DP-SGD mechanism $\mathcal{M}$, Loss function $\ell$, Number of observations $N$, Decision threshold $\tau$, Confidence level $\alpha$, Privacy parameter $\delta$ 
    
    \State \text{Initialize observations} $O, O'$
    \State \text{Craft Canary sample} $c$
    \State $D' \leftarrow D \cup \{c\}$

    \For{$i \in [N]$}
        \State $M_i \leftarrow  \mathcal{M}(D)$
        \State $M'_i \leftarrow  \mathcal{M}(D')$ 
    \EndFor

    \State \text{Craft Adversarial sample} $a$
    \For{$i \in [N]$}
        \State $O[i] \leftarrow \ell(M_i; a)$
        \State $O'[i] \leftarrow \ell(M'_i; a)$ 
    \EndFor

    \State \text{FPR} $\leftarrow \frac{1}{N} \sum_{o \in O} \mathbb{I}(o \geq \tau)$
    \State \text{FNR} $\leftarrow \frac{1}{N} \sum_{o \in O'} \mathbb{I}(o < \tau)$

    \State $\overline{\text{FPR}}
    \leftarrow \text{Clopper-Pearson(FPR,N,$\alpha$)}$
    \State $\overline{\text{FNR}}
    \leftarrow \text{Clopper-Pearson(FNR,N,$\alpha$)}$\
    
    \State $\varepsilon_{\text{emp}} \leftarrow \text{EstimateDP($\overline{\text{FPR}}$,$\overline{\text{FNR}}$,$\delta$)}$
    \State \Return $\varepsilon_{\text{emp}}$
    \end{algorithmic}
    \label{alg:auditing}
\end{algorithm}

%% file: main.bbl
\begin{thebibliography}{48}
\providecommand{\natexlab}[1]{#1}
\providecommand{\url}[1]{\texttt{#1}}
\expandafter\ifx\csname urlstyle\endcsname\relax
  \providecommand{\doi}[1]{doi: #1}\else
  \providecommand{\doi}{doi: \begingroup \urlstyle{rm}\Url}\fi

\bibitem[Abadi et~al.(2016)Abadi, Chu, Goodfellow, McMahan, Mironov, Talwar, and Zhang]{abadi2016deep}
Martin Abadi, Andy Chu, Ian Goodfellow, H~Brendan McMahan, Ilya Mironov, Kunal Talwar, and Li~Zhang.
\newblock Deep learning with differential privacy.
\newblock In \emph{Proceedings of the 2016 ACM SIGSAC conference on computer and communications security}, pages 308--318, 2016.

\bibitem[Aerni et~al.(2024)Aerni, Zhang, and Tram{\`e}r]{aerni2024evaluations}
Michael Aerni, Jie Zhang, and Florian Tram{\`e}r.
\newblock Evaluations of machine learning privacy defenses are misleading.
\newblock \emph{arXiv preprint arXiv:2404.17399}, 2024.

\bibitem[Altschuler and Talwar(2022)]{altschuler2022privacy}
Jason Altschuler and Kunal Talwar.
\newblock Privacy of noisy stochastic gradient descent: More iterations without more privacy loss.
\newblock \emph{Advances in Neural Information Processing Systems}, 35:\penalty0 3788--3800, 2022.

\bibitem[Andrew et~al.(2023)Andrew, Kairouz, Oh, Oprea, McMahan, and Suriyakumar]{andrew2023one}
Galen Andrew, Peter Kairouz, Sewoong Oh, Alina Oprea, H~Brendan McMahan, and Vinith Suriyakumar.
\newblock One-shot empirical privacy estimation for federated learning.
\newblock \emph{arXiv preprint arXiv:2302.03098}, 2023.

\bibitem[Annamalai(2024)]{annamalai2024s}
Meenatchi Sundaram Muthu~Selva Annamalai.
\newblock It's our loss: No privacy amplification for hidden state dp-sgd with non-convex loss.
\newblock \emph{arXiv preprint arXiv:2407.06496}, 2024.

\bibitem[Annamalai and De~Cristofaro(2024)]{annamalai2024nearly}
Meenatchi Sundaram Muthu~Selva Annamalai and Emiliano De~Cristofaro.
\newblock Nearly tight black-box auditing of differentially private machine learning.
\newblock \emph{arXiv preprint arXiv:2405.14106}, 2024.

\bibitem[Balle et~al.(2019)Balle, Barthe, Gaboardi, and Geumlek]{balle2019privacy}
Borja Balle, Gilles Barthe, Marco Gaboardi, and Joseph Geumlek.
\newblock Privacy amplification by mixing and diffusion mechanisms.
\newblock \emph{Advances in neural information processing systems}, 32, 2019.

\bibitem[Balle et~al.(2022)Balle, Cherubin, and Hayes]{balle2022reconstructing}
Borja Balle, Giovanni Cherubin, and Jamie Hayes.
\newblock Reconstructing training data with informed adversaries.
\newblock In \emph{2022 IEEE Symposium on Security and Privacy (SP)}, pages 1138--1156. IEEE, 2022.

\bibitem[Bassily et~al.(2014)Bassily, Smith, and Thakurta]{bassily2014private}
Raef Bassily, Adam Smith, and Abhradeep Thakurta.
\newblock Private empirical risk minimization: Efficient algorithms and tight error bounds.
\newblock In \emph{2014 IEEE 55th annual symposium on foundations of computer science}, pages 464--473. IEEE, 2014.

\bibitem[Bichsel et~al.(2021)Bichsel, Steffen, Bogunovic, and Vechev]{bichsel2021dp}
Benjamin Bichsel, Samuel Steffen, Ilija Bogunovic, and Martin Vechev.
\newblock Dp-sniper: Black-box discovery of differential privacy violations using classifiers.
\newblock In \emph{2021 IEEE Symposium on Security and Privacy (SP)}, pages 391--409. IEEE, 2021.

\bibitem[Bok et~al.(2024)Bok, Su, and Altschuler]{bok2024shifted}
Jinho Bok, Weijie Su, and Jason~M Altschuler.
\newblock Shifted interpolation for differential privacy.
\newblock \emph{arXiv preprint arXiv:2403.00278}, 2024.

\bibitem[Carlini et~al.(2019)Carlini, Liu, Erlingsson, Kos, and Song]{carlini2019secret}
Nicholas Carlini, Chang Liu, {\'U}lfar Erlingsson, Jernej Kos, and Dawn Song.
\newblock The secret sharer: Evaluating and testing unintended memorization in neural networks.
\newblock In \emph{28th USENIX security symposium (USENIX security 19)}, pages 267--284, 2019.

\bibitem[Carlini et~al.(2022)Carlini, Chien, Nasr, Song, Terzis, and Tramer]{carlini2022membership}
Nicholas Carlini, Steve Chien, Milad Nasr, Shuang Song, Andreas Terzis, and Florian Tramer.
\newblock Membership inference attacks from first principles.
\newblock In \emph{2022 IEEE Symposium on Security and Privacy (SP)}, pages 1897--1914. IEEE, 2022.

\bibitem[Cebere et~al.(2024)Cebere, Bellet, and Papernot]{cebere2024tighter}
Tudor Cebere, Aur{\'e}lien Bellet, and Nicolas Papernot.
\newblock Tighter privacy auditing of dp-sgd in the hidden state threat model.
\newblock \emph{arXiv preprint arXiv:2405.14457}, 2024.

\bibitem[Chadha et~al.(2024)Chadha, Jagielski, Papernot, Choquette-Choo, and Nasr]{chadha2024auditing}
Karan Chadha, Matthew Jagielski, Nicolas Papernot, Christopher Choquette-Choo, and Milad Nasr.
\newblock Auditing private prediction.
\newblock \emph{arXiv preprint arXiv:2402.09403}, 2024.

\bibitem[Chourasia et~al.(2021)Chourasia, Ye, and Shokri]{chourasia2021differential}
Rishav Chourasia, Jiayuan Ye, and Reza Shokri.
\newblock Differential privacy dynamics of langevin diffusion and noisy gradient descent.
\newblock \emph{Advances in Neural Information Processing Systems}, 34:\penalty0 14771--14781, 2021.

\bibitem[Clopper and Pearson(1934)]{clopper1934use}
Charles~J Clopper and Egon~S Pearson.
\newblock The use of confidence or fiducial limits illustrated in the case of the binomial.
\newblock \emph{Biometrika}, 26\penalty0 (4):\penalty0 404--413, 1934.

\bibitem[De et~al.(2022)De, Berrada, Hayes, Smith, and Balle]{de2022unlocking}
Soham De, Leonard Berrada, Jamie Hayes, Samuel~L Smith, and Borja Balle.
\newblock Unlocking high-accuracy differentially private image classification through scale.
\newblock \emph{arXiv preprint arXiv:2204.13650}, 2022.

\bibitem[Ding et~al.(2018)Ding, Wang, Wang, Zhang, and Kifer]{ding2018detecting}
Zeyu Ding, Yuxin Wang, Guanhong Wang, Danfeng Zhang, and Daniel Kifer.
\newblock Detecting violations of differential privacy.
\newblock In \emph{Proceedings of the 2018 ACM SIGSAC Conference on Computer and Communications Security}, pages 475--489, 2018.

\bibitem[Dong et~al.(2022)Dong, Roth, and Su]{dong2022gaussian}
Jinshuo Dong, Aaron Roth, and Weijie~J Su.
\newblock Gaussian differential privacy.
\newblock \emph{Journal of the Royal Statistical Society: Series B (Statistical Methodology)}, 84\penalty0 (1):\penalty0 3--37, 2022.

\bibitem[Doroshenko et~al.(2022)Doroshenko, Ghazi, Kamath, Kumar, and Manurangsi]{doroshenko2022connect}
Vadym Doroshenko, Badih Ghazi, Pritish Kamath, Ravi Kumar, and Pasin Manurangsi.
\newblock Connect the dots: Tighter discrete approximations of privacy loss distributions.
\newblock \emph{arXiv preprint arXiv:2207.04380}, 2022.

\bibitem[Dwork et~al.(2006)Dwork, McSherry, Nissim, and Smith]{dwork2006calibrating}
Cynthia Dwork, Frank McSherry, Kobbi Nissim, and Adam Smith.
\newblock Calibrating noise to sensitivity in private data analysis.
\newblock In \emph{Theory of Cryptography: Third Theory of Cryptography Conference, TCC 2006, New York, NY, USA, March 4-7, 2006. Proceedings 3}, pages 265--284. Springer, 2006.

\bibitem[Feldman et~al.(2018)Feldman, Mironov, Talwar, and Thakurta]{feldman2018privacy}
Vitaly Feldman, Ilya Mironov, Kunal Talwar, and Abhradeep Thakurta.
\newblock Privacy amplification by iteration.
\newblock In \emph{2018 IEEE 59th Annual Symposium on Foundations of Computer Science (FOCS)}, pages 521--532. IEEE, 2018.

\bibitem[Gopi et~al.(2021)Gopi, Lee, and Wutschitz]{gopi2021numerical}
Sivakanth Gopi, Yin~Tat Lee, and Lukas Wutschitz.
\newblock Numerical composition of differential privacy.
\newblock \emph{Advances in Neural Information Processing Systems}, 34:\penalty0 11631--11642, 2021.

\bibitem[Haim et~al.(2022)Haim, Vardi, Yehudai, Shamir, and Irani]{haim2022reconstructing}
Niv Haim, Gal Vardi, Gilad Yehudai, Ohad Shamir, and Michal Irani.
\newblock Reconstructing training data from trained neural networks.
\newblock \emph{Advances in Neural Information Processing Systems}, 35:\penalty0 22911--22924, 2022.

\bibitem[Hayes et~al.(2017)Hayes, Melis, Danezis, and De~Cristofaro]{hayes2017logan}
Jamie Hayes, Luca Melis, George Danezis, and Emiliano De~Cristofaro.
\newblock Logan: Membership inference attacks against generative models.
\newblock \emph{arXiv preprint arXiv:1705.07663}, 2017.

\bibitem[Jagielski et~al.(2020)Jagielski, Ullman, and Oprea]{jagielski2020auditing}
Matthew Jagielski, Jonathan Ullman, and Alina Oprea.
\newblock Auditing differentially private machine learning: How private is private sgd?
\newblock \emph{Advances in Neural Information Processing Systems}, 33:\penalty0 22205--22216, 2020.

\bibitem[Jeon et~al.(2024)Jeon, Jeung, Kim, No, and Choi]{jeon2024information}
Dongjae Jeon, Wonje Jeung, Taeheon Kim, Albert No, and Jonghyun Choi.
\newblock An information theoretic metric for evaluating unlearning models.
\newblock \emph{arXiv preprint arXiv:2405.17878}, 2024.

\bibitem[Kairouz et~al.(2015)Kairouz, Oh, and Viswanath]{kairouz2015composition}
Peter Kairouz, Sewoong Oh, and Pramod Viswanath.
\newblock The composition theorem for differential privacy.
\newblock In \emph{International conference on machine learning}, pages 1376--1385. PMLR, 2015.

\bibitem[Koskela et~al.(2020)Koskela, J{\"a}lk{\"o}, and Honkela]{koskela2020computing}
Antti Koskela, Joonas J{\"a}lk{\"o}, and Antti Honkela.
\newblock Computing tight differential privacy guarantees using fft.
\newblock In \emph{International Conference on Artificial Intelligence and Statistics}, pages 2560--2569. PMLR, 2020.

\bibitem[LeCun et~al.(1998)LeCun, Cortez, and Burges]{lecun1998mnist}
Y.~LeCun, C.~Cortez, and C.~C. Burges.
\newblock The mnist database of handwritten digits, 1998.
\newblock URL \url{http://yann.lecun.com/exdb/mnist/}.

\bibitem[Maddock et~al.(2022)Maddock, Sablayrolles, and Stock]{maddock2022canife}
Samuel Maddock, Alexandre Sablayrolles, and Pierre Stock.
\newblock Canife: Crafting canaries for empirical privacy measurement in federated learning.
\newblock \emph{arXiv preprint arXiv:2210.02912}, 2022.

\bibitem[Mahloujifar et~al.(2024)Mahloujifar, Melis, and Chaudhuri]{mahloujifar2024auditing}
Saeed Mahloujifar, Luca Melis, and Kamalika Chaudhuri.
\newblock Auditing $ f $-differential privacy in one run.
\newblock \emph{arXiv preprint arXiv:2410.22235}, 2024.

\bibitem[Mironov(2017)]{mironov2017renyi}
Ilya Mironov.
\newblock R{\'e}nyi differential privacy.
\newblock In \emph{2017 IEEE 30th computer security foundations symposium (CSF)}, pages 263--275. IEEE, 2017.

\bibitem[Nasr et~al.(2021)Nasr, Songi, Thakurta, Papernot, and Carlin]{nasr2021adversary}
Milad Nasr, Shuang Songi, Abhradeep Thakurta, Nicolas Papernot, and Nicholas Carlin.
\newblock Adversary instantiation: Lower bounds for differentially private machine learning.
\newblock In \emph{2021 IEEE Symposium on security and privacy (SP)}, pages 866--882. IEEE, 2021.

\bibitem[Nasr et~al.(2023)Nasr, Hayes, Steinke, Balle, Tram{\`e}r, Jagielski, Carlini, and Terzis]{nasr2023tight}
Milad Nasr, Jamie Hayes, Thomas Steinke, Borja Balle, Florian Tram{\`e}r, Matthew Jagielski, Nicholas Carlini, and Andreas Terzis.
\newblock Tight auditing of differentially private machine learning.
\newblock In \emph{32nd USENIX Security Symposium (USENIX Security 23)}, pages 1631--1648, 2023.

\bibitem[Papernot et~al.(2016)Papernot, Abadi, Erlingsson, Goodfellow, and Talwar]{papernot2016semi}
Nicolas Papernot, Mart{\'\i}n Abadi, Ulfar Erlingsson, Ian Goodfellow, and Kunal Talwar.
\newblock Semi-supervised knowledge transfer for deep learning from private training data.
\newblock \emph{arXiv preprint arXiv:1610.05755}, 2016.

\bibitem[Shokri et~al.(2017)Shokri, Stronati, Song, and Shmatikov]{shokri2017membership}
Reza Shokri, Marco Stronati, Congzheng Song, and Vitaly Shmatikov.
\newblock Membership inference attacks against machine learning models.
\newblock In \emph{2017 IEEE symposium on security and privacy (SP)}, pages 3--18. IEEE, 2017.

\bibitem[Stadler et~al.(2022)Stadler, Oprisanu, and Troncoso]{stadler2022synthetic}
Theresa Stadler, Bristena Oprisanu, and Carmela Troncoso.
\newblock Synthetic data--anonymisation groundhog day.
\newblock In \emph{31st USENIX Security Symposium (USENIX Security 22)}, pages 1451--1468, 2022.

\bibitem[Steinke et~al.(2024{\natexlab{a}})Steinke, Nasr, Ganesh, Balle, Choquette-Choo, Jagielski, Hayes, Thakurta, Smith, and Terzis]{steinke2024last}
Thomas Steinke, Milad Nasr, Arun Ganesh, Borja Balle, Christopher~A Choquette-Choo, Matthew Jagielski, Jamie Hayes, Abhradeep~Guha Thakurta, Adam Smith, and Andreas Terzis.
\newblock The last iterate advantage: Empirical auditing and principled heuristic analysis of differentially private sgd.
\newblock \emph{arXiv preprint arXiv:2410.06186}, 2024{\natexlab{a}}.

\bibitem[Steinke et~al.(2024{\natexlab{b}})Steinke, Nasr, and Jagielski]{steinke2024privacy}
Thomas Steinke, Milad Nasr, and Matthew Jagielski.
\newblock Privacy auditing with one (1) training run.
\newblock \emph{Advances in Neural Information Processing Systems}, 36, 2024{\natexlab{b}}.

\bibitem[Thaker et~al.(2024)Thaker, Hu, Kale, Maurya, Wu, and Smith]{thaker2024position}
Pratiksha Thaker, Shengyuan Hu, Neil Kale, Yash Maurya, Zhiwei~Steven Wu, and Virginia Smith.
\newblock Position: Llm unlearning benchmarks are weak measures of progress.
\newblock \emph{arXiv preprint arXiv:2410.02879}, 2024.

\bibitem[Tramer et~al.(2022)Tramer, Terzis, Steinke, Song, Jagielski, and Carlini]{tramer2022debugging}
Florian Tramer, Andreas Terzis, Thomas Steinke, Shuang Song, Matthew Jagielski, and Nicholas Carlini.
\newblock Debugging differential privacy: A case study for privacy auditing.
\newblock \emph{arXiv preprint arXiv:2202.12219}, 2022.

\bibitem[Ye and Shokri(2022)]{ye2022differentially}
Jiayuan Ye and Reza Shokri.
\newblock Differentially private learning needs hidden state (or much faster convergence).
\newblock \emph{Advances in Neural Information Processing Systems}, 35:\penalty0 703--715, 2022.

\bibitem[Yeom et~al.(2018)Yeom, Giacomelli, Fredrikson, and Jha]{yeom2018privacy}
Samuel Yeom, Irene Giacomelli, Matt Fredrikson, and Somesh Jha.
\newblock Privacy risk in machine learning: Analyzing the connection to overfitting.
\newblock In \emph{2018 IEEE 31st computer security foundations symposium (CSF)}, pages 268--282. IEEE, 2018.

\bibitem[Yousefpour et~al.(2021)Yousefpour, Shilov, Sablayrolles, Testuggine, Prasad, Malek, Nguyen, Ghosh, Bharadwaj, Zhao, et~al.]{yousefpour2021opacus}
Ashkan Yousefpour, Igor Shilov, Alexandre Sablayrolles, Davide Testuggine, Karthik Prasad, Mani Malek, John Nguyen, Sayan Ghosh, Akash Bharadwaj, Jessica Zhao, et~al.
\newblock Opacus: User-friendly differential privacy library in pytorch.
\newblock \emph{arXiv preprint arXiv:2109.12298}, 2021.

\bibitem[Zanella-Beguelin et~al.(2023)Zanella-Beguelin, Wutschitz, Tople, Salem, R{\"u}hle, Paverd, Naseri, K{\"o}pf, and Jones]{zanella2023bayesian}
Santiago Zanella-Beguelin, Lukas Wutschitz, Shruti Tople, Ahmed Salem, Victor R{\"u}hle, Andrew Paverd, Mohammad Naseri, Boris K{\"o}pf, and Daniel Jones.
\newblock Bayesian estimation of differential privacy.
\newblock In \emph{International Conference on Machine Learning}, pages 40624--40636. PMLR, 2023.

\bibitem[Zhu et~al.(2020)Zhu, Yu, Chandraker, and Wang]{zhu2020private}
Yuqing Zhu, Xiang Yu, Manmohan Chandraker, and Yu-Xiang Wang.
\newblock Private-knn: Practical differential privacy for computer vision.
\newblock In \emph{Proceedings of the IEEE/CVF Conference on Computer Vision and Pattern Recognition}, pages 11854--11862, 2020.

\end{thebibliography}
